# Experiment-based model of Er/Yb gain medium for fiber amplifiers and lasers

D.K. Vysokikh,[1,2] A.P. Bazakutsa,[3] A.V. Dorofeenko,[1,2,3,4] and O.V. Butov[1,4]

[1]Moscow Institute of Physics and Technology, Dolgoprudny, Moscow region, Russia

[2]N.L. Dukhov Moscow Institute of Autimatics (VNIIA), Moscow, Russia

[3]Institute for Radio-engineering and Electronics RAS, Moscow, Russia

[4]Institute for Theoretical and Applied Electromagnetics RAS, Moscow, Russia

A model of an $Er^{3+}/Yb^{3+}$ co-doped glass gain medium is considered. The system of rate equations taking into account a minimum set of processes required to describe experimental data is built. As a result, a moderate number of fitting parameters is used, thereby increasing the model reliability. For the populations of the energy levels of $Er^{3+}$ and $Yb^{3+}$ ions, analytic expressions are obtained, as well as for the gain factor of the medium. The model was validated by a series of measurements of gain in $Er^{3+}/Yb^{3+}$ fiber amplifiers of different lengths at variable pump powers. The parameters of the model are found to fit the experimental data, taking into account the spatial inhomogeneity of the pump power along the fiber. The maximum population inversion of erbium at high pumping powers is analytically expressed, which describes the "bottleneck" effect. The power required to pump the fiber along the entire length is found.

## 1. Introduction

Highly efficient excitation scheme of gain medium based on $Er^{3+}$ and $Yb^{3+}$ co-doped quartz glass is widely used in fiber lasers and amplifiers [1], in microdisk lasers [2] etc.

As a rule, the Er/Yb scheme is described by the rate equations model [3-6]. In Ref. [3], a basic model of an Er/Yb-based laser is proposed without taking into account the absorption saturation of ytterbium ions, and a correspondence of the calculated lasing curves with the experimental ones was obtained. The development of this model, which taked into account the saturation of the populations of the $Yb^{3+}$ excited state, as well as the absorption of pump power by $Er^{3+}$ ions, up-conversion processes, excited state absorption, excitation transfer between the $Yb^{3+}$ and $Er^{3+}$ ions with an absorption of the second photon, was carried out in Refs. [3-5, 7]. In some versions of the model, the presence of $Yb^{3+}$ ions, which do not interact with $Er^{3+}$ ions, was additionally taken into account [4]. The resulting model made it possible to calculate the spontaneous emission spectra and the excitation relaxation rates [8], find the dependence of the fiber amplifier gain on the concentration of $Yb^{3+}$ ions, pump power and fiber length [9], determine the lasing threshold [5], analyze the possible values of the laser quantum yield [10]. The model parameters were found from the best correspondence between the experimental and calculated data for the luminescence kinetics curves [4], photoluminescence intensity [8], lifetime of excited energy levels [9], lasing intensity [5].

In the Er/Yb gain medium, an effect of so-called "bottleneck" is known, which consists in impossibility to increase the $Er^{3+}$ ion excited state population above some finite value even at uncreasingly high pump power. Under this condition, induced up and down transitions in $Yb^{3+}$ ions dominate other processes. The ratio of the up and down transition rates determines the population of the $Yb^{3+}$ excited state, which in turn governs the rate of excitation transfer to the $Er^{3+}$ ion. As a result, the $Er^{3+}$ ion excited level population does not exceed some finite value below unity. In the present paper, this value is found analytically.

The models proposed in the aforementioned papers make it possible to investigate the dependence of the properties of the Er/Yb gain medium on various parameters. It should be noted that the dependence of these properties on the concentrations of $Er^{3+}$ and $Yb^{3+}$ ions is nontrivial, because the excitation transfer rate from $Yb^{3+}$ to $Er^{3+}$ depends on the distance between the ions, which, in turn, depends in a non-obvious manner on the concentrations due to formation of effective field for $Er^{3+}$ by $Yb^{3+}$ ions [11], due to clustering effects, etc. In some papers, an assumption was made on the linear dependence of the excitation transfer rate on the $Yb^{3+}$ concentration [5, 8, 12]. On the other hand, we determine the model parameters from measurements carried out at a fixed ion concentrations, but vary the pump power and fiber length, which allows us to remove the described uncertainty.

In this paper, we consider a model based on a system of rate equations that takes into account only the effects necessary to describe the experiment, which made it possible to minimize the number of fitting parameters and thereby increase the model reliability. An analytical solution was obtained for the amplification factor in active fiber. The "bottleneck" effect was studied and the corresponding maximum population inversions of $Er^{3+}$ and $Yb^{3+}$ were found. An experiment was carried out, which made it possible to validate the model and determine its parameters.

**2. Experimental methods**

A pump-probe experiments were carried out to measure the gain of active fiber sections. An experimental scheme is presented in Fig. 1. Pump (at a wavelength of 976 nm) and probe signals were injected through a micro-optical fiber multiplexer 7 into the Yb and Er co-doped fiber section 8. The source of pump radiation was a semiconductor laser diode 1 up to 400 mW with a fiber output. A CW semiconductor laser 3 with a fiber output served as the source of the probing signal, which power was confined at a –33.5 dBm level by an optical attenuator 5. The wavelength of the probing signal source was tuned in the range of 1526 to 1566 nm. After the active fiber, an output fused fiber optic multiplexer 9 was placed. The spectrum of the output signal was measured by an optical spectrum analyzer Yokogawa AQ6370D 12, the power of unabsorbed pump radiation was recorded with an optical power meter JDSU OLP-85 11. In the experiment, the amplified signal spectrum was studied by the breaking method [7]. We studied 3, 4 and 5 cm long active fiber samples.

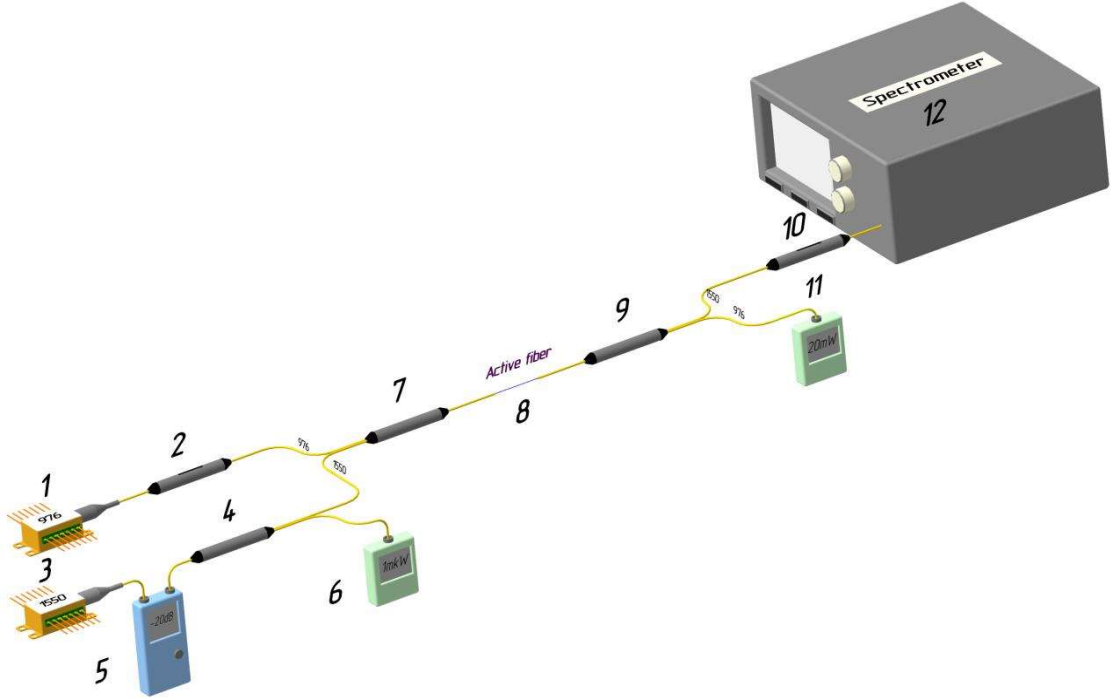

Fig. 1. Experimental scheme. 1 – 976 nm semiconductor laser, 2 – optical isolator at 976 nm, 3 – 1550 nm semiconductor laser, 4 – fused fiber splitter with a division factor of 1/99, 5 – tunable fiber attenuator, 6 –1550 nm (input signal) power meter, 7 – microoptical fiber multiplexer for 1550 and 976 nm, 8 – active fiber sample under study, 9 – fused fiber demultiplexer for wavelengths of 1550 and 976 nm, 10 – 1550 nm fiber insulator, 11 – power meter of unabsorbed pump radiation at 976 nm, 12 – optical spectrum analyzer with fiber input.

**3. Model of Er/Yb gain medium with the pump power inhomogeneous along the fiber**

Consider an Er/Yb active fiber. Let us describe the $Yb^{3+}$ and $Er^{3+}$ ions interaction with local pump field, and restrict ourselves to pair interactions. Consider the following processes (Fig. 2): absorption and stimulated emission in $Yb^{3+}$ ion at the pump frequency, which are characterized by $\gamma_{p1}$ and $\gamma_{p2}$ rates, absorption and stimulated emission of the $Er^{3+}$ ion at the signal frequency at the $\Gamma_{i1}$ and $\Gamma_{i2}$ rates, relaxation of excitation of $Yb^{3+}$ и $Er^{3+}$ ions at $\gamma_1$ и $\Gamma_1$ rates, and nonradiative energy transfer from $Yb^{3+}$ to $Er^{3+}$ ion at a $\gamma_T$ rate. Number of excitation quanta per unit fiber length transmitted from all ytterbium ions per unit time, equals $C_{Yb}\gamma_T$, with $C_{Yb}$ being the $Yb^{3+}$ ions concentration per unit fiber length. The same number of excitations is received by erbium ions: $C_{Er}\Gamma_T = C_{Yb}\gamma_T$. Thus, due to the difference in $C_{Er}$ and $C_{Yb}$, excitation transfer rate, $\gamma_T$, from $Yb^{3+}$ ions and excitation reception rate, $\Gamma_T$, by $Er^{3+}$ ions, are different.

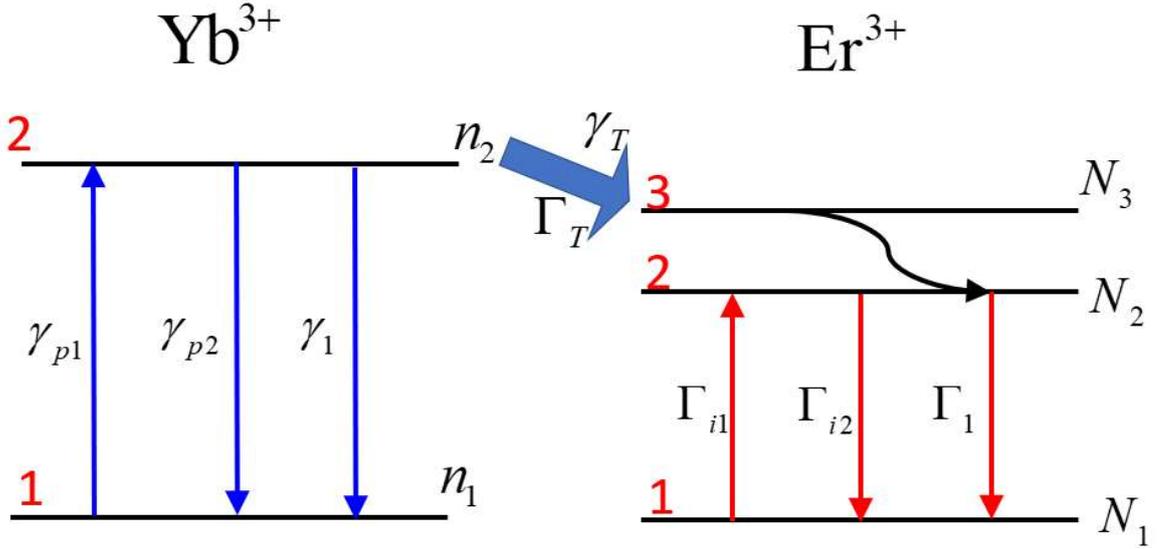

Fig. 2. Energy levels in Er/Yb system and the processes taken into account

With these processes taken into account, the rate equations are written as

$$\dot{n}_2 = \gamma_{p1} n_1 - \gamma_{p2} n_2 - \gamma_1 n_2 - \gamma_T n_2 N_1, \qquad (1)$$

$$\dot{N}_1 = \Gamma_1 N_2 - \Gamma_T n_2 N_1 - \Gamma_{i1} N_1 + \Gamma_{i2} N_2, \qquad (2)$$

where the probability of excitation transfer depends on both the population of the upper ytterbium level and the availability of the upper (i.e., population of the lower) erbium level. Note that, in this model, we neglect the possibility of an induced transition of Er from level 1 to level 3, since the Er absorption cross section is approximately 10 times smaller than the Yb absorption cross section, and the Yb concentration is at least several times higher. The validity of this approximation is discussed in Ref. [3].

Let us neglect induced processes in $Er^{3+}$ ions (see some discussion in Conclusions):

$$\dot{n}_2 = \gamma_{p1} n_1 - \gamma_{p2} n_2 - \gamma_1 n_2 - \gamma_T n_2 N_1, \qquad (3)$$

$$\dot{N}_1 = \Gamma_1 N_2 - \Gamma_T n_2 N_1. \qquad (4)$$

In the stationary case, we have $\dot{N}_1 = 0$ and $\dot{n}_2 = 0$. Solving together Eqs. (3) and (4), we obtain for $N_1$

$$N_1 = -\frac{1}{2}\left( \frac{\gamma_1 + \gamma_{p1} + \gamma_{p2}}{\gamma_T} - 1 + \frac{\gamma_{p1}}{\gamma_T}\frac{\Gamma_T}{\Gamma_1} + \right.$$

$$\left. + \sqrt{4\frac{\gamma_{p1}}{\gamma_T}\frac{\Gamma_T}{\Gamma_1}\left(\frac{\gamma_1 + \gamma_{p1} + \gamma_{p2}}{\gamma_T}\right) + \left(\left(\frac{\gamma_1 + \gamma_{p1} + \gamma_{p2}}{\gamma_T} + 1\right) - \frac{\gamma_{p1}}{\gamma_T}\frac{\Gamma_T}{\Gamma_1}\right)^2}\right) \qquad (5)$$

After designating

$$\xi = \frac{\gamma_1 + \gamma_{p1} + \gamma_{p2}}{\gamma_T}, \quad \theta = \frac{\Gamma_T}{\Gamma_1} \quad \text{and} \quad \phi = \frac{\gamma_{p1}}{\gamma_T}\frac{\Gamma_T}{\Gamma_1}, \tag{6}$$

we get a simplified expression:

$$N_1 = \left(1 - \xi - \phi + \sqrt{(\xi + \phi)^2 + 2\xi - 2\phi + 1}\right)/2. \tag{7}$$

Similarly, we obtain an expression for $n_2$:

$$n_2 = \frac{\sqrt{(\xi + \phi)^2 + 2\xi - 2\phi + 1} - \xi + \phi - 1}{2\xi\theta}. \tag{8}$$

Next, it is necessary to separate the dependences of the populations on the pump power in an explicit form, so we define the variables:

$$\chi_1 = \frac{\gamma_{p1}}{\gamma_T P_p}, \quad \chi_2 = \frac{\gamma_{p2}}{\gamma_T P_p}, \quad g = \frac{\gamma_1}{\gamma_T}. \tag{9}$$

Note that the pumping rates $\gamma_{p1}$ and $\gamma_{p2}$ are proportional to the pump power $P_p$. Therefore, the value of $\chi_1$ does not depend on $P_p$. With an account of Eq. (6), we get explicit dependencies of $\xi$ and $\phi$ on the pump power:

$$\xi = g + (\chi_1 + \chi_2)P_p, \quad \phi = \chi_1 \theta P_p. \tag{10}$$

Substituting (10) into (7) and (8), we obtain the functions of the model parameters, $N_1(P_p, \chi_1, \chi_2, g, \theta)$ and $n_2(P_p, \chi_1, \chi_2, g, \theta)$, which will be used in further calculations.

Note that the value of (7) is in the range from 0 to 1 for any value $\xi > 0$ and $\phi > 0$, which corresponds to the physical meaning of the population $N_1$. However, the value of (8) can exceed 1 for small $\theta$. This is because the parameterization (6) does not take into account the relationship between parameters. This relationship is taken into account by Eq. (10), so that the resulting dependences $N_1(P_p, \chi_1, \chi_2, g, \theta)$ and $n_2(P_p, \chi_1, \chi_2, g, \theta)$ are in the range from 0 to 1.

Let us proceed to the description of the pump power distribution along the fiber. The absorption coefficient of pump power, which can be expressed as the imaginary part of the wave number $\beta_p''$ of the fiber mode at the pump frequency, depends on the populations of the ytterbium levels. To find this dependence, we write the differential equation for the mode energy flux, $P_p$:

$$dP_p / dz = -2\beta_p'' P_p. \tag{11}$$

Here, the factor "2" is due to the fact that the wave number describes the spatial dependence of the fields, while the power attenuation/gain is described by twice the imaginary part of the wave number.

On the other hand, the energy transferred to a unit length of the gain medium per unit time is equal to $dP_p/dz = C_{Yb}\hbar\omega_p(\gamma_{p1}n_1 - \gamma_{p2}n_2)$. As a result, we arrive at the relation

$$\beta_p'' = \frac{C_{Yb}\hbar\omega_p}{2}\frac{\gamma_{p1}n_1 - \gamma_{p2}n_2}{P_p}. \tag{12}$$

We also introduce the notation for the linear part of the absorption coefficient of the pump field, which corresponds to small field strengths at which $n_1 = 1$, $n_2 = 0$:

$$\beta_{0p}'' = \frac{C_{Yb}\hbar\omega_p}{2}\frac{\gamma_{p1}}{P_p}. \tag{13}$$

Then, Eq. (12) is rewritten in the form

$$\beta_p'' = \beta_{0p}''\left(1 - \frac{\gamma_{p2} + \gamma_{p1}}{\gamma_{p1}}n_2\right). \tag{14}$$

Let us now consider the field at the frequency of the working transition of $Er^{3+}$ ($N_2 \to N_1$), which obeys the equation

$$dP/dz = -2\beta''P. \tag{15}$$

For the imaginary part of the wave number, $\beta''$, of the fiber mode at the transition frequency, in analogy to (12), we arrive at the ratio

$$\beta'' = \beta_0''\left(\frac{\Gamma_{i1} + \Gamma_{i2}}{\Gamma_{i1}}N_1 - \frac{\Gamma_{i2}}{\Gamma_{i1}}\right), \tag{16}$$

where $\beta_0''$ is a linear part of the field absorption coefficient at the transition frequency. Let us define a new variable $\chi_3 = \Gamma_{i2}/\Gamma_{i1}$.

With Eqs. (7), (8) with the account of the relations (6), (10) we find the populations of the levels, which allow to solve the nonlinear equation (11) for the pump power distribution.

**4. Determination of model parameters from experimental data**

Let us determine the model parameters from the experimental data. We present the results of the experiment on the output probe signal power measurement as an average attenuation coefficient

$$\langle\beta''\rangle = \frac{1}{2L}\ln\left(\frac{P(0)}{P(L)}\right), \tag{17}$$

where $L$ is the fiber section length. Here, the coefficient «2» in the denominator [see also the remark after Eq. (11)] is due to the spatial dependence of the radiation power, which is characterized by a doubled wavenumber.

Let us find the same value from the model calculation. By integrating Eq. (15), we write $\ln\left(\frac{P(0)}{P(L)}\right) = 2\int_0^L \beta''(z)dz$. Thus,

$$\langle\beta''\rangle = \frac{1}{L}\int_0^L \beta''(z)dz. \tag{18}$$

In order to express $\beta''(z)$ in terms of the model parameters, we numerically solve the differential equation (11) with a boundary condition $P_p(0) = P_{p0}$, i.e., with the given input pump power. Here, we use $\beta_p''$ in the form (14), where $n_2$ depends on $\xi$, $\phi$ and $\theta$ [see (8)], which in turn depend on $P_p$ and on the model parameters $\chi_1$, $\chi_2$, $g$ and $\theta$ [see Eq. (10)]. In this way, we obtain the distribution of the pump power $P_p$ along the fiber. Further, by calculating the population $N_1$ distribution along the fiber [see Eq. (7)], we get $\langle\beta''\rangle$ [see (16), (18)], which depends on the input pump power $P_{p0}$ and on the model parameters $\chi_1$, $\chi_2$, $\chi_3$, $g$, $\theta$, $\beta''_{0p}$, $\beta''$. Note that the use of Eq. (17) would require an additional calculation of the probe signal power distribution $P(z)$ along the fiber, which made us using Eq. (18).

By varying the model parameters, we obtain their optimal values from the best fit (by the least squares method) of the calculated and experimental dependences of $\langle\beta''\rangle$ on the pump power for three sections (3, 4 and 5 cm long) of the fiber (Fig. 3).

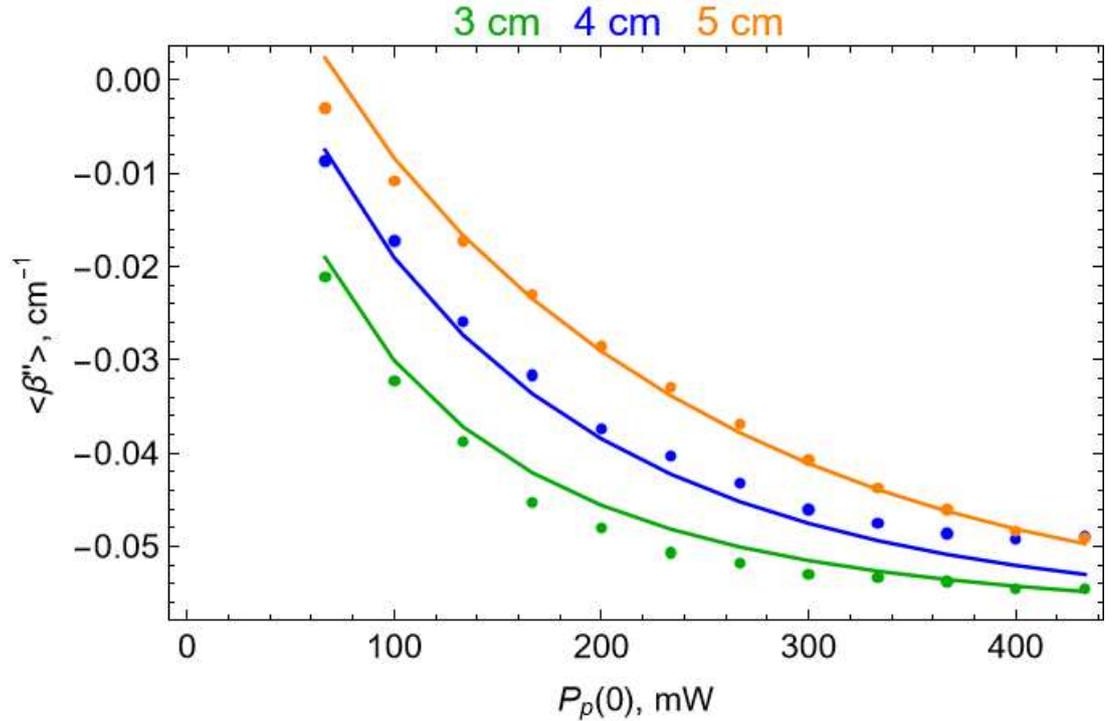

Fig. 3. Experimental (points) and calculated (curves) dependences of the average attenuation coefficient of the probe signal on the pump power for fiber sections of 3 cm (green), 4 cm (blue) and 5 cm long (orange). The resulting parameters' values: $g = 58.9$, $\chi_1 = 0.29$, $\chi_2 = 0.01$, $\chi_3 = 1.21$, $\theta = 20.05$, $\beta_0'' = 0.054$, $\beta_{0p}'' = 0.49$.

### 5. Numerical and analytical results

Let us study the properties of the gain medium based on the obtained computational model.

Let us first consider the populations of ytterbium and erbium ions against the local pump energy density, in which these ions are immersed. An increase in the pump power leads to an increase in the populations of the upper levels of ytterbium and erbium (Fig. 4). The properties of ytterbium as absorbing or amplifying medium at the pump frequency is determined by the sign of a quantity $\gamma_{p2} n_2 - \gamma_{p1} n_1$ [see Eq. (3)], which may be rewritten as $\frac{\gamma_{p1} + \gamma_{p2}}{2} \delta n$. Here,

$$\delta n = \frac{2\gamma_{p2}}{\gamma_{p1} + \gamma_{p2}} n_2 - \frac{2\gamma_{p1}}{\gamma_{p1} + \gamma_{p2}} n_1 \qquad (19)$$

is an effective population inversion, the negative sign of which indicates that the ytterbium ions are absorbing at the pump frequency. However, the erbium population inversion $\delta N = \frac{2\Gamma_{i2}}{\Gamma_{i1} + \Gamma_{i2}} N_2 - \frac{2\Gamma_{i1}}{\Gamma_{i1} + \Gamma_{i2}} N_1$ becomes positive at the pump power exceeding a certain threshold value, and the medium as a whole is amplifying at the frequency of the probe signal.

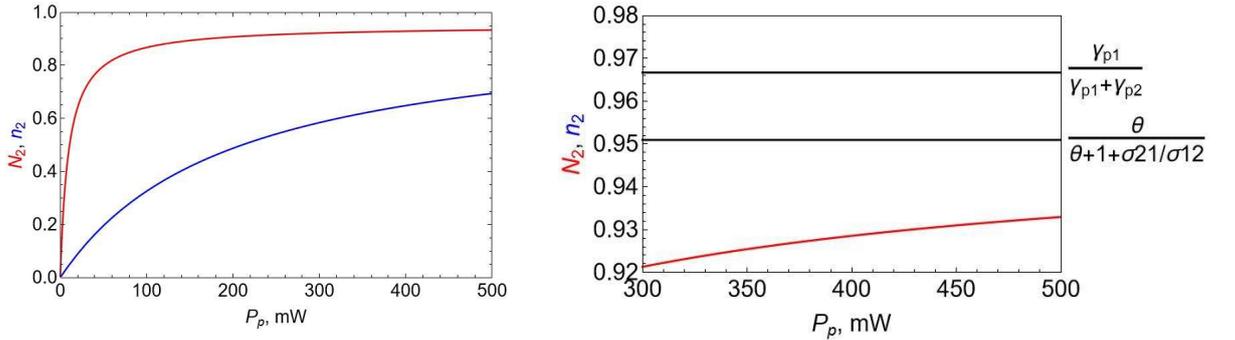

Fig. 4. Populations of the excited states of the erbium ($N_2$) and ytterbium ($n_2$) ions depending on the local power $P_p$ of the pump field at the parameters' values determined from the experiment (Section 4). The figure is shown in two scales. A characteristic value of $n_2$, $\frac{\gamma_{p1}}{\gamma_{p1} + \gamma_{p2}}$, below which the ytterbium ion is absorbing, is shown. Also shown is a maximal value of $N_2$, $\frac{\theta}{\theta + 1 + \gamma_{p2}/\gamma_{p1}}$, corresponding to an infinitely high pump power.

Let us consider the question of the maximum value of the erbium population inversion, which arises in the limit of high pump powers. It is known that [4, 9], the population of the second level of erbium ions does not tend to unity with an increasing pump power (the "bottleneck" effect). In our model, a limit of $\gamma_{p1}, \gamma_{p2} \to \infty$ leads to $N_1 \to \dfrac{1+\gamma_{p2}/\gamma_{p1}}{1+\gamma_{p2}/\gamma_{p1}+\theta}$, i.e., $N_2 \to \dfrac{\theta}{\theta+1+\gamma_{p2}/\gamma_{p1}}$ (Appendix), and to the population inversion $\delta N = \dfrac{2}{\Gamma_{i1}+\Gamma_{i2}} \dfrac{\Gamma_{i2}\theta - \Gamma_{i1}(1+\gamma_{p2}/\gamma_{p1})}{1+\gamma_{p2}/\gamma_{p1}+\theta}$. At the chosen wavelength, the transition upwards in the ytterbium ion is much more efficient than the transition downwards: $\gamma_{p2}/\gamma_{p1} \ll 1$. Furthermore, $\theta \gg 1$, meaning that the interaction of ions is faster than the relaxation of the erbium ion. In these approximations, the population inversion of erbium approaches the maximal value $\delta N = \dfrac{2\Gamma_{i2}}{\Gamma_{i1}+\Gamma_{i2}}$. Thus, the choice of parameters allows you to level the effect of the "bottleneck" (Fig. 4).

Let us consider a distribution of the pump power and of the absorption/amplification coefficients at the pump and signal frequencies (Fig. 5). Note that at large enough $z$ the dependence of $P_p(z)$ is exponential, which is due to the independence of the absorption coefficient of the field strength (linear region). At small $z$, the pump radiation power is high, and the medium exhibits nonlinear properties: the absorption coefficient decreases with an increasing field. Particularly, in this region, the pump radiation power is characterized not by an exponential but by a linear spatial dependence (Fig. 5a).

In accordance with the foresaid, in the linear region, the absorption coefficients tend to stationary values $\beta_0''$ и $\beta_{0p}''$ (Fig. 5b), which correspond to the ions being in the ground state. In the nonlinear region, the values of $\beta''$ и $\beta_p''$ decrease with respect to the stationary values, in particular, $\beta''$ becomes negative, which corresponds to amplification of the signal.

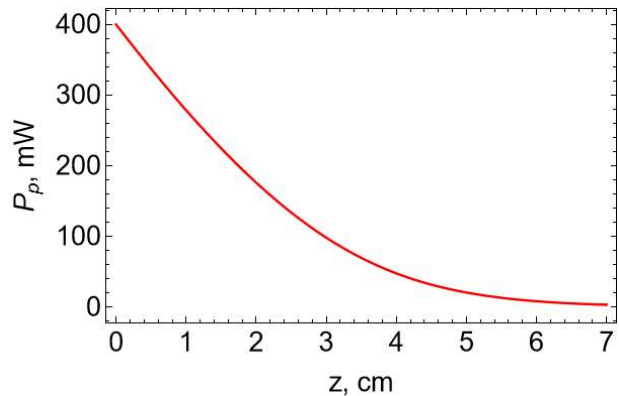

(a)

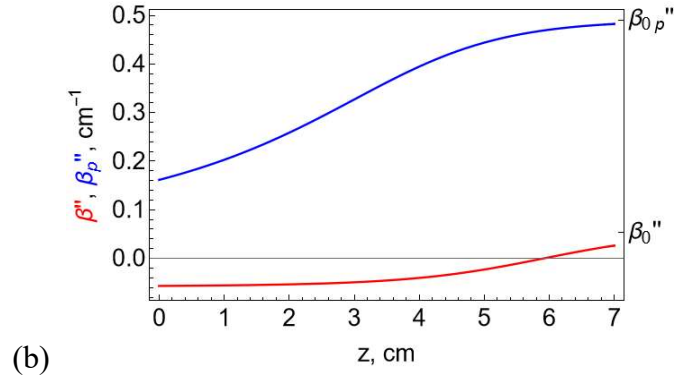

(b)

Fig. 5. Spatial distribution of the pump power (a) and of coefficients of absorption/amplification (b) of pump radiation (blue curve) and of the signal (red curve).

Let us estimate the level of pump radiation power, above which the medium becomes amplifying, i.e., $\delta N = 0$. Corresponding population is $N_1 \triangleq N_1^{ap} \sim \dfrac{\Gamma_{i2}}{\Gamma_{i1}+\Gamma_{i2}}$. From Eq. (7) follows that $N_1 = \left(1-\xi-\phi+\sqrt{(\xi+\phi)^2+2\xi-2\phi+1}\right)/2$ and

$$\xi = N_1^2 + N_1(\xi+\phi-1), \qquad (20)$$

leading to

$$\xi = N_1 \frac{N_1+\phi-1}{1-N_1}. \qquad (21)$$

As a result, we obtain from (9) и (10):

$$\frac{\gamma_1+\gamma_{p1}+\gamma_{p2}}{\gamma_T} = N_1 \frac{N_1 + \dfrac{\gamma_{p1}}{\gamma_T}\dfrac{\Gamma_T}{\Gamma_1}-1}{1-N_1}, \qquad (22)$$

which, after a substitution of the value of $\gamma_1$, $\gamma_{p1}$, $\gamma_{p2}$ expressed through the model parameters at $N_1 = N_1^{ap}$ leads to

$$P_P^* = \frac{\left(1-\dfrac{\chi_3}{1+\chi_3}\right)g + \dfrac{\chi_3}{1+\chi_3} - \left(\dfrac{\chi_3}{1+\chi_3}\right)^2}{\dfrac{\chi_3}{1+\chi_3}\chi_1\theta - \left(1-\dfrac{\chi_3}{1+\chi_3}\right)(\chi_1+\chi_2)}, \qquad (23)$$

or, after transformations,

$$P_P^* = \frac{g + \chi_3 - \dfrac{\chi_3^2}{1+\chi_3}}{\chi_3\chi_1\theta - \chi_1 - \chi_2}. \qquad (24)$$

If the local pump power exceeds this value, then the medium becomes amplifying. At our parameters, $P_p^* \approx 8.8$ mW. By injecting such power to the fiber input, $P_{p0} = P_p^*$, we get a short amplifying section at the beginning of the fiber. In order to obtain amplification in the entire fiber of the length $L$, it is necessary to apply a larger input power, determined by the condition $P_p(L) \geq P_p^*$ (Fig. 6).

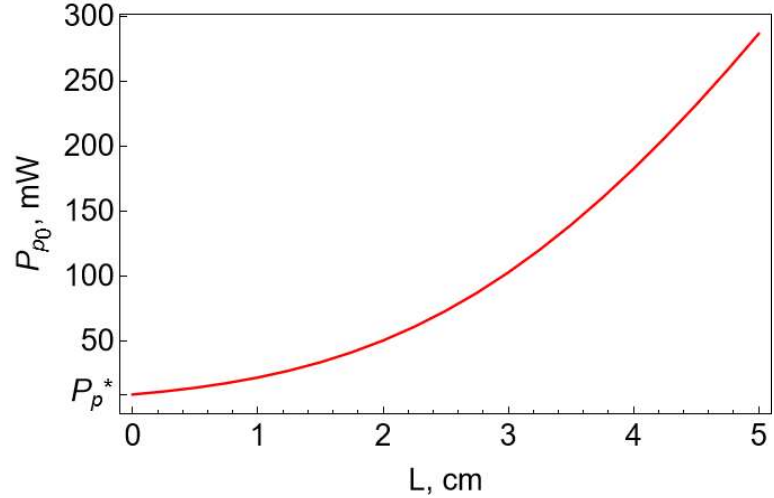

Fig. 6. Dependence of the input pump power required for complete pumping of the fiber on its length.

**6. Conclusions**

In this paper, we present a rate equations based model of a fiber amplifier co-doped by Er/Yb ions, which takes into account only those processes that are necessary to describe the gain medium. This simplification made it possible to limit the number of fitting parameters and thus ensure the model reliability. These parameters were found by fitting the numerically built curves to experimental data for the dependence of the signal gain, and a good agreement was obtained. The distributions of the pump and useful gains along the fiber length were calculated. The "bottleneck" effect was studied and the threshold values of Er and Yb population inversions were determined.

The proposed model can be used to describe optical amplifiers, as well as fiber lasers with a Bragg cavity. In such devices, pump power inhomogeneity along the length of the fiber is significant due to the absorption of the pump field. To describe lasing, Eqs. (1) and (2) may be used, which takes induced transitions in the erbium ion into account. However, this is not always necessary. For example, to determine the lasing threshold, Eqs. (3) and (4) is enough, since the generation is determined by the instability of the state with a low field intensity [5, 13].

**Funding**

This research was funded by Russian Science Foundation, project number 20-72-10057.

**Appendix**

Let us find a maximal value of the population $N_1$ achieved at large pump power. With the notations

$$A = \xi + \phi, \; B = 2(\xi - \phi) \qquad (25)$$

we rewrite Eq. (7) in the form

$$N_1 = (1 - A + \sqrt{A^2 + B + 1}). \qquad (26)$$

Next, we expand the square root in a Taylor series in terms of a small parameter $\dfrac{1}{A^2 + B}$:

$$N_1 = \frac{1}{2}\left(1 - A + \sqrt{A^2 + B}\left(1 + \frac{1}{2(A^2 + B)} + ...\right)\right) \qquad (27)$$

Then, in the denominator in (27), we take out $A^2$ and expand by the parameter $\frac{B}{A^2}$. We obtain

$$N_1 = \frac{1}{2}\left(1 - A + \sqrt{A^2 + B}\left(1 + \frac{1}{2A^2\left(1 + \frac{B}{A^2}\right)} + ...\right)\right) = \frac{1}{2}\left(1 + \frac{B}{2A} + ...\right). \qquad (28)$$

Thus, in the limit of high pump power, we find

$$\lim_{\substack{\gamma_{p1} \to \infty \\ \gamma_{p2} \to \infty}} N_1 = \frac{1}{2}\left(1 + \frac{B}{2A}\right), \qquad (29)$$

or, with the account of (25),

$$\lim_{\substack{\gamma_{p1} \to \infty \\ \gamma_{p2} \to \infty}} N_1 = \frac{1}{1 + \phi/\xi}. \qquad (30)$$

Further, in the same limit, $\phi/\xi = \frac{\gamma_{p1}}{\gamma_1 + \gamma_{p1} + \gamma_{p2}} \frac{\Gamma_T}{\Gamma_1} \to \frac{\theta}{1 + \gamma_{p2}/\gamma_{p1}}$. Therefore,

$$\lim_{\substack{\gamma_{p1} \to \infty \\ \gamma_{p2} \to \infty}} N_1 = \frac{1 + \gamma_{p2}/\gamma_{p1}}{1 + \gamma_{p2}/\gamma_{p1} + \theta}. \qquad (31)$$